# Chronological Citation Recommendation with Time Preference

Shutian Ma[1], Heng Zhang[1], Chengzhi Zhang[1,*], Xiaozhong Liu[2]

[1] Department of Information Management, Nanjing University of Science and Technology, Nanjing, China, 210094
[2] Department of Information & Library Science, Indiana University Bloomington, USA,
mashutian0608@hotmail.com, zh_heng@njust.edu.cn, zhangcz@njust.edu.cn,
liu237@indiana.edu,

**Abstract.** Citation recommendation is an important task to assist scholars in finding candidate literature to cite. Traditional studies focus on static models of recommending citations, which do not explicitly distinguish differences between papers that are caused by temporal variations. Although, some researchers have investigated chronological citation recommendation by adding time related function or modeling textual topics dynamically. These solutions can hardly cope with function generalization or cold-start problems when there is no information for user profiling or there are isolated papers never being cited. With the rise and fall of science paradigms, scientific topics tend to change and evolve over time. People would have the time preference when citing papers, since most of the theoretical basis exist in classical readings that published in old time, while new techniques are proposed in more recent papers. To explore chronological citation recommendation, this paper wants to predict the time preference based on user queries, which is a probability distribution of citing papers published in different time slices. Then, we use this time preference to re-rank the initial citation list obtained by content-based filtering. Experimental results demonstrate that task performance can be further enhanced by time preference and it's flexible to be added in other citation recommendation frameworks.

## 1    Introduction

Due to the increasing of scientific publication amount, citation recommendation has arisen researchers' attention, which aims to help people find appropriate and necessary work to cite based on the given user queries like keywords, short text or existing reference list. So far, there are three main tracks of strategy in the citation recommendation systems (Khusro, Ali, & Ullah, 2016): content-based filtering (Arif, 2016), collaborative filtering (Hernando, Bobadilla, & Ortega, 2016) and hybrid ones (H. Liu et al., 2015). Since papers published in different time slices are processed with static models, most studies have ignored chronological nature of citation recommendation. As it is described in Jiang, Liu, and Gao (2015), when users are retrieving papers, their information need may feature different variations depending on varying time slices. Proper citations can not only support claims in one's work, but also give insight into the early period of modern science (Bornmann & Mutz, 2015). Normally, when scholars are doing literature survey, he/she is likely to review papers among different time slices to follow and display the development of research topic. As Figure 1 shows, in an example of the application scenario which is to recommend a citation paper list for scholars based on user query, if a user's query is related to "*Topic Modeling*", he/she might be mostly interested in publications focusing on "*Latent Dirichlet Allocation*" in 2003. When there is a need to investigate earlier publications, "*Latent Semantic Indexing*" or "*Probabilistic Latent Semantic Indexing*" between 1990 and 2000, and some other studies published before 1990 will also have probabilities to be cited in the literature review. Therefore, there would be a time preference when scholars are citing papers due to the topic evolution or users' information need (classical readings VS. recent advances). Since the scientific topics are changing and evolving over time, time factor should play an important role to influence or even infer the citation ranking list indirectly.

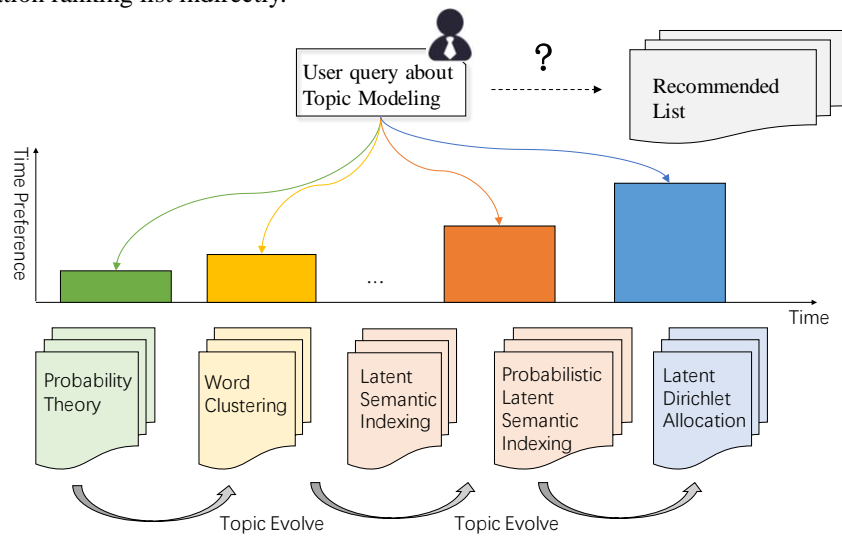

Figure 1: Diagrammatic sketch of time preference for citing

To the best of our knowledge, only a few prior efforts have attempted to conduct chronological citation recommendation so far. Some researchers established functions to measure the change of user interests over citation published time (Mahdabi & Crestani, 2014; Wu, Hua, Li, & Pei, 2012). However, this kind of method might not be universal to all scientific papers from various domains. Firstly, scientists' attention often differs over topics. Secondly, the development processes in different disciplines are various from each other. Recently, chronological dynamic topic models are also proposed to recommend time-series ranking lists (Jiang, Liu, & Gao, 2014; Jiang et al.,

2015) while some time-related features have to be constructed based on prior knowledge and experiences. Lately, by parameterizing the age distribution of a work's references (number of years a reference has been published for), researchers found that the age distribution of prior knowledge is particularly linked to tomorrow's breakthroughs (Mukherjee, Romero, Jones, & Uzzi, 2017). Inspired by information need shifting (Jiang et al., 2015) and the age distribution of reference list, we want to model the time preference to do chronological citation recommendations. Since research topics are evolving all over the time, people would have different preferences over the papers published in different time periods. In this paper, we convert this time preference into probability distribution that people cite over different time periods. First of all, multi-layer perceptron model is used to predict time preference. Then, probability of citing the candidate citation is obtained according to time preference and publishing time of the given candidate citation. Finally, the new recommended list will be generated based on a weighted similarity between candidate citation and user query using that probability. To sum up, time preference will be applied for re-ranking candidate citations which are obtained by content-based filtering.

The remainder of this paper is organized as follows. Section 2 provides a brief review of traditional citation recommendation, chronological citation recommendation and neural network-based citation recommendation. Section 3 elaborates detailed information about the proposed model. Experiment and results are illustrated in Section 4. Finally, conclusion, limitations and future work are demonstrated in Section 5.

## 2      Related Work

We summarize the related work of this paper and arrange it from three dimensions. First, we will go through the relevant researches about traditional citation recommendation. Second, we consider methods that applied in chronological citation recommendation. Finally, several neural network-based studies will be introduced.

### 2.1    Traditional Citation Recommendation

Given the user query and set of published papers, citation recommendation is a task of recommending suitable citations to scholars based on their queries, which can help them improve searching efficiency on papers and prevent missing important paper citations (Ma, Liu, & Zhang, 2020). Slightly different from some personalized paper recommenders which provide articles that users might be interested in, citation recommendation is more concerned with recommending papers that should be referred to (Huang et al., 2012; Huang, Wu, Liang, Mitra, & Giles, 2015). Different from studies trying to encode users' preference based on user profiling (Hong, Jeon, & Jeon, 2013), citation recommenders make more use of users' textual input, like keywords (Wu et al., 2012), part of reference list (El-Arini & Guestrin, 2011) or short text (Dai, Zhu, Wang, & Carley, 2019; Tang & Zhang, 2009), and try to identify the hidden evidence behind the huge published literature. More specifically, textual content (He, Pei, Kifer, Mitra, & Giles, 2010), bibliographic networks (Cai, Han, & Yang, 2018; Chen, Liu, Zhao, &

Zhang, 2019; X. Liu, Yu, Guo, & Sun, 2014) and meta-data (Sun et al., 2018) contained in scientific repository are all mined for generation, ranking and evaluation of recommended citations.

When utilizing textual content of user query and candidate citations, user query has also evolved into a format which is context, referring to a bag of words (He et al. 2010). He et al (2010) depict two types of context-aware citation recommendations, global recommendation and local recommendation, which are mentioned as context-based citation recommendation as well. In global recommendation, users can submit a manuscript or few sentences as the query. The global context is title and abstract of the source paper. The local context is text surrounding a citation or placeholder. In local recommendation, users will provide a local context with respect to the source paper. Generally, these two kinds of citation recommendations are making full utilization of content information, such as user query and textual content of papers. So far, researchers have tried to improve citation generation performance from aspects of text representation, text similarity measurement and classifier selections . Kazemi and Abhari (2017) compared TF-IDF model with popular word-embedding to see which one works better in content-based filtering. In work by Sun et al. (Sun et al., 2013) , three types of similarity measures are integrated: keyword similarity, journal similarity, and author similarity to measure relevance of the articles to researchers. He et al. (He, Kifer, Pei, Mitra, & Giles, 2011) compared four different models to examine the relevance between user query and document corpus. Ensemble of decision trees shows better performance than language model, contextual similarity and topical relevance. Combinations between similarity computation algorithms (cosine similarity, Euclidean distance, Jaccard coefficient, Pearson correlation coefficient) and classifiers (Random Forest, Recursive Partitioning and Boosted tree) are also tested by Magara, Ojo, and Zuva (2018). There are also researchers who are trying to analyze user queries. Färber, Thiemann, and Jatowt (2018) proposed a system which can determine for each potential citation context, if it is actually "cite-worthy."

With more bibliographic data open to people, network-based citation recommenders are proposed by researchers. Aiming at finding important or neighbor paper nodes, multi-layer networks are built by paper, author, venue and keyword nodes. Graph related algorithms like random walk or PageRank are applied or optimized for finding candidate citations. Meng, Gao, Li, Sun, and Hou (2013) build a four-layer graph based on authors, papers, topics and words, apply Random walk-based algorithm for recommendation. Cai, Han, Li, et al. (2018) project paper, author and venue into a three-layered graph and build a mutual reinforcement model with three PageRank-like models for each of them. Ranking of each node is derived not only from the relationship within itself but also affected by the other two kinds of nodes. Another graph-based model integrating fine-grained co-authorship as well as author–paper, paper–citation, and paper–keyword relations is designed by Guo et al. (2017). To modify relationship of co-authors, they define a collaboration influence distribution in different topics as authors' collaboration features.

As it shows, most traditional work in citation recommendations has been geared towards static recommendation framework (He et al., 2011; He et al., 2010; Sun et al., 2013; Tang & Zhang, 2009), where the text-based and relation-based approaches ignore the chronological nature.

## 2.2 Chronological Citation Recommendation

There are also few attempts to take time factor into considerations in citation recommendation. Jiang et al. proposed that the researcher's information need may change over time, as the scientific contents of a specific research topic are evolving with time (Jiang et al., 2014). In order to characterize such shifting and intrinsic citation time-decay, they integrated text search and citation link navigation in a chronological dynamic topic model environment to recommend time-series ranking lists based on users' initial textual information needs. In the following work, they employed meta-path to add the possibility of enriching semantic information between papers in the citation networks (Jiang et al., 2015). Except building models to learn time factor, few studies modified time information to be a weighting function and this has been tried in only limited ways until now. When ranking candidate citations based on various evidences, Wu et al. suggest that authors prefer recently published citations to support and illustrate the novelty of their own works (Wu et al., 2012). Therefore, they directly set different likelihoods of a paper being cited in different years and assume that when a paper has been published more than 20 years, it received no attention. Mahdabi et al. proposed a patent citation recommendation model which employs time-aware random walk on the weighted patent citation network (Mahdabi & Crestani, 2014). The authors used a time granularity to modify the initial probability for selecting a node that discounts the node according to its age.

Above studies have figured out to integrate the time factor into procedure of citation recommendations. However, there do exist limitations in the generalization of methods that adding time information into functions. Complexity of algorithms that conducting dynamic chronological feature extraction needs further optimizations as well.

## 3 Methodology

Our citation recommendation model contains two main steps. In the first step, we will encode the user query and learns probability distribution for citing preference over time slices. The second step will take advantage of the predicted time reference to re-rank candidate citations list obtained from content-based filtering for generating the final recommended list.

### 3.1 Citing Time Preference Prediction

In the first step, we want to predict citing time preference based on user query and existed publication dataset using neural network model. To encapsulate some hidden patterns over both textual topics and citation behaviors, we feed content embedding and latent citation embedding into the multi-layer perception model. The diagrammatic sketch is given in Figure 2.

### 3.1.1 Generation of model input

First of all, given the query collection $Q$ and published paper collection $D$, vector-based representations of query and paper are firstly learned over the whole document set via a pre-selected content embedding model. For a given user query $x$, its content embedding $x^{content}$ is then obtained.

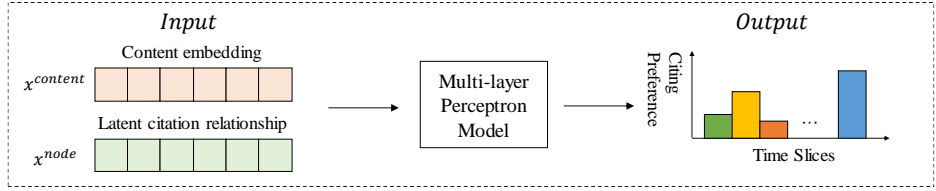

Figure 2: Diagrammatic sketch of generating model input

Since papers with similar topics tend to share similar citations, we enrich query information from the aspect of latent citation relations. We use the citation graph constructed by $D$ to learn representation of each paper via a pre-selected node embedding model. Then, as Figure 3 shows, the pre-trained content embeddings are utilized to find out query's $k$ nearest neighbor papers in $D$ which are published in the same time period as the query based on cosine similarity (Mihalcea, Corley, & Strapparava, 2006):

$$Cosine\ Similarity = \frac{\sum_{i=1}^{m}(q_i \times t_i)}{\sqrt{\sum_{i=1}^{m}(q_i)^2} \times \sqrt{\sum_{i=1}^{m}(t_i)^2}} \quad (1)$$

Where, $(q_1, q_2, \cdots, q_m)$ is the content embedding of query, $(t_1, t_2, \cdots, t_m)$ is the content embedding of papers in $D$ and $m$ is the parameter of vector dimension. After finding out the $k$ nearest neighbors, average pooling is conducted on these papers' node embeddings to generate a vector that is made to imply the latent citation relationship of user query.

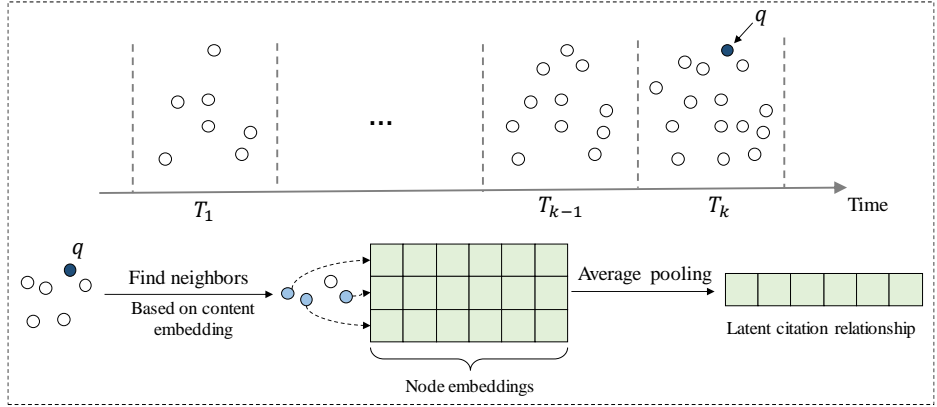

Figure 3: Guideline of generating latent citation relationship

Then, we will have the input layer for multi-layer perceptron (MLP) model (Chaudhuri & Bhattacharya, 2000). $x^{content}$ is the content embedding of user query and $x^{node}$ is the latent citation relationship. Since we integrate user's textual query and citation networks of query's neighbor paper. It enables recommendation in a cold-start environment. There is no need to prepare initial set of citations relevant to the users' research topics or previous citing records. Isolated papers with no citation connections will still have a chance to be recommended.

### 3.1.2 Network structure of MLP model

Figure 3 displays the structure of multi-layer perceptron model built for predicting time preference.

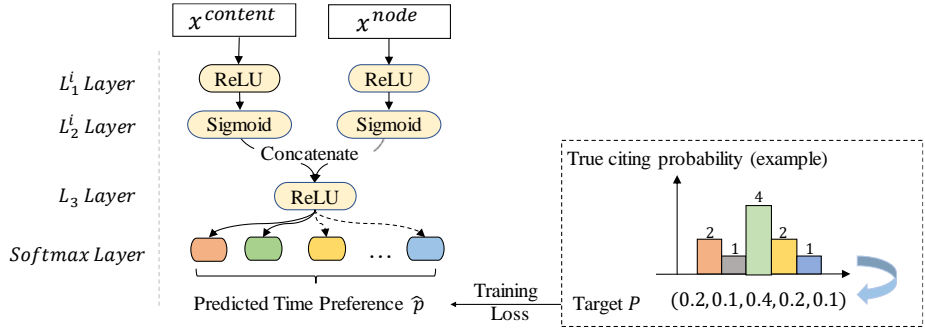

Figure 4: Network structure of multi-layer perceptron model

As it is shown, the input layer is followed with two separate and fully connected hidden layers, $L_1^i$ and $L_2^i$, where $i \in \{content, node\}$ denotes relevant input vectors. Dimensional parameters for different layers are set without tuning. Dimension of layer $L_1^i$ is set to be 70, dimension of layer $L_2^i$ is set to be 50. These two layers are defined as following:

$$L_1^i = ReLU(W_1 x^i + b_1) \qquad (2)$$
$$L_2^i = Sigmoid(W_2 x^i + b_2) \qquad (3)$$

$W_1, W_2$ and $b_1, b_2$ denote related weights and biases, respectively. In the third hidden layer, we concatenate the two feature vectors into one, dimension of layer $L_3$ is set to be 80. Probability of predicted time preference $\hat{P}$ is finally achieved by applying a softmax function over the last fully connected layer:

$$L_3 = ReLU(W_3 [L_2^{content}, L_2^{node}] + b_3) \qquad (4)$$
$$\hat{P} = softmax(W_4 L_3) \qquad (5)$$

$W_3$ and $b_3$ denote related weights and biases, respectively. Here, $\hat{P} = (\hat{p}_1, \hat{p}_2, \cdots, \hat{p}_t)$ and $t$ refers to the number of time slices, $\hat{p}_i$ refers to the predicted probability of the $i_{th}$ time slice. For each user query, $P = (p_1, p_2, \cdots, p_t)$ and $p_i$ is the true citing probability modified by the normalized paper citation times of the $i_{th}$ time slice, which is calculated as:

$$p_i = \frac{Count_i}{Count_1 + Count_2 + \cdots + Count_t} \tag{6}$$

Where, $Count_i$ is the number of cited papers that published in the $i_{th}$ time slice. For example, if a paper cites 10 reference papers in total and these papers are published in five time slices as it's shown in Figure 3. The true citing probability of this citing paper is $(0.2, 0.1, 0.4, 0.2, 0.1)$. During model training, cross entropy between true citing probability and predicted probability is chosen to be loss function and its formula is given below:

$$Loss(P, \hat{P}) = -\sum_{i=0}^{|P|} p_i \log \hat{p}_i \tag{7}$$

Besides, we use Adam algorithm (Kingma & Ba, 2014) for efficient stochastic optimization and learning rate is set to be 0.001.

### 3.2 Candidate Citation Generation

When a new query $q$ with its time information come, we firstly find the query's neighbor papers based on content similarity. Then we can have the node embedding of given query inferred from the average node embeddings of neighbor nodes. Based on our trained MLP model, we obtain the time preference $(\hat{p}_1, \hat{p}_2, \cdots, \hat{p}_t)$ of this query. In order to generate an initial citation set, given the published paper collection $D = (d_1, d_2, \cdots, d_n)$, similarity between the query and target paper will be calculated over all pairs of query and papers, i.e., $S(q, d_1), S(q, d_2), \cdots, S(q, d_n)$, where we denote with $S$ the cosine similarity computed via content embeddings. Finally, we use the predicted time preference of query $q$ to be the weighting factor for each target paper $d$ according to paper's published time period. For instance, if $d$ is published in the $i_{th}$ time period, the final utility for it to be cited by user query $q$ is computed as follows:

$$S(q, d)_{weighted} = S(q, d) \times \frac{1}{1 + e^{-\hat{p}_i}} \tag{8}$$

Here, sigmoid function is applied to modify predicted time preference. Then, our model will recommend citation list of papers ranked from largest to least value in terms of $S(q, d)_{weighted}$.

## 4 Experimental Analysis

### 4.1 Experimental Setup

This section will introduce the experimental data, evaluation metrics, model parameter setting, and the baselines applied in this paper.

#### 4.1.1 Experimental data

In this paper, we applied two different datasets for experiments: PubMed and DBLP. For PubMed data, we extracted 544, 511 publications with 916, 860 citation relations from PubMed Central[1] between 1877 and 2013 to be our initial data set. We split the corpus into seven time slices: pre-1995, 1996-2000, 2001-2003, 2004-2005, 2006-2007, 2008-2009, and 2010-2013. Each time slice contains a similar number of papers shown in Table 1. Then, a collection of 12, 882 papers are obtained firstly, whose whole reference data (abstract and published year) can be found from the original data set. Furthermore, each of these papers has over 30 references which are published in more than 5 time slices. Except the citation relationships between citing paper and cited paper, we also have the citing counts for each citation relationship. For MLP training and testing purposes, we randomly divide the entire collection into two disjoint sets, 11, 882 paper with their references are deemed as training set to train the multi-layer perception model and the remaining papers fall into the test set (1000 papers with their references).

Table 1 Paper numbers of different time slices of PubMed dataset

| Time Slice | Paper Number | Time Slice | Paper Number |
|---|---|---|---|
| pre-1995 | 68, 523 | 2006-2007 | 82, 868 |
| 1996-2000 | 77, 290 | 2008-2009 | 90, 165 |
| 2001-2003 | 77, 504 | 2010-2013 | 77, 954 |
| 2004-2005 | 70, 207 | | |

For DBLP data, we obtained the DBLP-Citation-network V12 collection from Aminer[2]. we extracted 2, 217, 689 publications that published before 2013 to be our initial data set. We split the corpus into eight time slices: pre-1995, 1996-2000, 2001-2003, 2004-2005, 2006-2007, 2008-2009, 2010-2011 and 2012-2013. Each time slice contains a similar number of papers shown in Table 2.

Table 2 Paper numbers of different time slices of DBLP dataset

| Time Slice | Paper Number | Time Slice | Paper Number |
|---|---|---|---|
| pre-1995 | 209, 784 | 2006-2007 | 274, 456 |
| 1996-2000 | 195, 106 | 2008-2009 | 327, 508 |
| 2001-2003 | 196, 118 | 2010-2011 | 374, 769 |
| 2004-2005 | 208, 282 | 2012-2013 | 431, 666 |

Then, a collection of 8700 papers are obtained firstly, whose whole reference data (abstract and published year) can be found from the original data set. Furthermore, each of these papers has over 30 references which are published in more than 5 time slices. For MLP training and testing purposes, we randomly divide the entire collection into two disjoint sets, 7900 paper with their references are deemed as training set to train the multi-layer perception model and the remaining papers fall into the test set (800 papers with their references).

**4.1.2 Evaluation metrics and parameters**

Specifically, abstracts of testing papers are used as the user queries and a paper's reference list is treated as the ground truth. Moreover, the cited times of reference papers indicates the importance of each paper (for NDCG relevance score). Since the largest reference number in test data is 391, we will recommend a citation list containing 500 papers for each user query from 57, 464 papers. Several evaluation metrics are applied here[3]: mean average precision (MAP) (Voorhees, 2000) and normalized discounted cumulative gain (NDCG) (Yilmaz, Kanoulas, & Aslam, 2008), mean reciprocal rank (MRR) (Craswell, 2009), Precision@N and Recall@N. MAP is often used as a single summary evaluation statistic. The mean average precision for a submission consisting of multiple topics is mean of average precision scores of each topic in submission. Given $|Q|$ queries, it's calculated as follows:

$$MAP = \frac{\sum_{q \in Q} Average\ Precision}{|Q|} \quad (9)$$

---
[1] Available at: https://www.ncbi.nlm.nih.gov/pmc/
[2] Available at: https://www.aminer.cn/citation

NDCG is usually truncated at a particular rank level (e.g. the first 10 retrieved documents) to emphasize importance of the first retrieved documents. DCG is a weighted sum of relevancy degree of ranked items. NDCG normalizes DCG, which is simply DCG measure of the best ranking result. Suppose IDCG is DCG measure of the true ranking result, formula of NDCG is shown below:

$$NDCG = \frac{DCG}{IDCG} \quad (10)$$

Mean reciprocal rank is inverse position of the first relevant document in result and is well-suited to applications in which only the first result matters. Its formula is given here:

$$MRR = \frac{\sum_{q \in Q} Reciprocal\ Rank}{|Q|} \quad (11)$$

When learning content embedding and node embedding, we use Doc2Vec[4] and Node2Vec[5] respectively based on the abstract content and citation network without citation links between test data. Doc2Vec (Le & Mikolov, 2014) is an extension algorithm to word2vec (Mikolov, Chen, Corrado, & Dean, 2013) to learn document-level embeddings. Node2Vec (Grover & Leskovec, 2016) is an algorithmic framework to learn continuous feature representations for nodes in networks (return parameter $p$ and in-out parameter $q$ are all set to be 0.25). The two embedding sizes are all set to be 100. When selecting query's $k$ nearest neighbor papers, we test the $k$ with the following values: 10, 15, 20, 50, 100 to 200. When receiving content embedding and latent citation relationship in the input layer, we scale each feature by its maximum absolute value[6].

**4.1.3 Baselines for comparisons**

In order to evaluate the quality of our proposed model, we select five different approaches to be the baselines. First of all, the traditional CBF algorithm is conducted which only depends on the cosine similarity computed using Doc2Vec results. The second algorithm is Dynamic Topic Model[7], which can capture the evolution of topics in a sequentially organized corpus of documents (Blei & Lafferty, 2006). Here we set the topic number for each document to be 10 without parameter tuning. Referring to weighted content similarity, three different weight functions are applied. The first is CiteRank algorithm[8] which refers to the PageRank over citation data (Walker, Xie, Yan, & Maslov, 2007). Second is publication time preference mentioned in (Wu et al., 2012). This paper establishes a piecewise function to measure the change of user interesting over the citation published time:

$$preference(c) = \begin{cases} \sigma^7 & Y_t = Y_c \\ \sigma^{Y_t - Y_c - 1} & 1 \leq Y_t - Y_c \leq 20 \\ \sigma^{20} & Y_t - Y_c > 20 \end{cases} \quad (12)$$

---

[3] Available at: https://github.com/usnistgov/trec_eval
[4] Available at: https://radimrehurek.com/gensim/models/doc2vec.html
[5] Available at: https://github.com/thunlp/OpenNE
[6] Available at: https://scikit-learn.org/stable/modules/preprocessing.html

Where, $Y_t$ is the publication year of target paper and $Y_c$ is the publication year of a candidate citation paper. σ is the damping factor to model user preference over times. Generally, setting σ as [0.8~0.9] is enough and we set it to be 0.8 in this paper. The last one is document freshness in time aware network (Mahdabi & Crestani, 2014). The authors modify the initial probability to assign importance to newer nodes as described in the following equation:

$$\rho_i = e^{\frac{-age}{\tau_d}} \quad (13)$$

Except the baseline methods relevant with time information, we also tried a model called WHIN-CSL proposed by Chen et al. (2019). They made use of weighted heterogeneous information network to obtain feature representation of each vertex. Then, references are recommended through linear combination among multimodal similarities. Basically, WHIN-CSL compute three views of similarities: abstract-abstract similarity $\mu_1(cp_c, tp_t)$, paper-paper vertex similarity $\mu_2(cp_c, tp_t)$ and paper-author vertexes similarity $\mu_3(cp_c, tp_t)$. For each target paper $tp_t = (ab_t, au_t)$, the task is to recommend a list of papers from candidate paper set $CP$. The conditional probability $P_r(cp_c|tp_t)$ is computed via equation below:

$$P_r(cp_c|tp_t) = w_1\mu_1(cp_c, tp_t) + w_2\mu_2(cp_c, tp_t) + (1 - w_1 - w_2)\mu_3(cp_c, tp_t) \quad (14)$$

Where, $w_1 + w_2 < 1$. In this paper, the PubMed data doesn't contain author information, so we only make use of the abstract-abstract similarity and paper-paper vertex similarity to generate the probability. Parameter setting of $w_1$ and $w_2$ will be given in experimental results.

### 4.2 Result Analysis

#### 4.2.1 Performance between baselines and our model

---

[7] Available at: https://radimrehurek.com/gensim/models/ldaseqmodel.html
[8] Available at: https://github.com/elifesciences/citerank

Table 3 Evaluation metrics of recommendation methods using PubMed dataset

| Metrics / Method | MAP | NDCG | MMR | P@30 | R@30 |
|---|---|---|---|---|---|
| CBF | 0.0772 | 0.2680 | 0.6152 | 0.1757 | 0.0968 |
| DTM | 0.0090 | 0.0798 | 0.1175 | 0.0307 | 0.0162 |
| Citerank | 0.0526 | 0.2208 | 0.2678 | 0.1387 | 0.0783 |
| Freshness | 0.0284 | 0.1794 | 0.2439 | 0.0593 | 0.0307 |
| Preference | 0.0396 | 0.2126 | 0.4473 | 0.0878 | 0.0477 |
| WHIN-CSL_1 | 0.0007 | 0.0109 | 0.0293 | 0.0037 | 0.0020 |
| WHIN-CSL_2 | 0.0029 | 0.0302 | 0.1019 | 0.0144 | 0.0079 |
| WHIN-CSL_3 | 0.0159 | 0.0907 | 0.3254 | 0.0516 | 0.0279 |
| WHIN-CSL_4 | 0.0891 | 0.2514 | 0.6819 | 0.1840 | 0.0977 |
| Our model | 0.0779 | 0.2694 | 0.6167 | 0.1766 | 0.0974 |

| Metrics / Method | MAP@30 | MAP@100 | NDCG@30 | NDCG@100 |
|---|---|---|---|---|
| CBF | 0.0499 | 0.0648 | 0.1835 | 0.2071 |
| DTM | 0.0034 | 0.0053 | 0.0215 | 0.0798 |
| Citerank | 0.0249 | 0.0393 | 0.1177 | 0.1554 |
| Freshness | 0.0099 | 0.0147 | 0.0486 | 0.0772 |
| Preference | 0.0204 | 0.0259 | 0.0978 | 0.1190 |
| WHIN-CSL_1 | 0.0005 | 0.0005 | 0.0025 | 0.0040 |
| WHIN-CSL_2 | 0.0020 | 0.0023 | 0.0103 | 0.0144 |
| WHIN-CSL_3 | 0.0101 | 0.0121 | 0.0412 | 0.0507 |
| WHIN-CSL_4 | 0.0500 | 0.0671 | 0.1335 | 0.1623 |
| Our model | 0.0505 | 0.0653 | 0.1859 | 0.2081 |

Table 3 shows evaluation metrics using PubMed dataset. For the model of WHIN-CSL, since the author information are unavailable, conditional probability only contains abstract-abstract similarity and paper-paper vertex similarity. For WHIN-CSL_1, $w_1$ is set to be 0.9 and $w_2$ is set to be 0.1. For WHIN-CSL_2, $w_1$ is set to be 0.8 and $w_2$ is set to be 0.2. For WHIN-CSL_3, $w_1$ is set to be 0.7 and $w_2$ is set to be 0.3. For WHIN-CSL_4, $w_1$ is set to be 0.6 and $w_2$ is set to be 0.4. Experimental results of our model are obtained when $k = 100$, where k is paper number of query's nearest neighbors and MAP obtains the highest value at this point. It can be observed that, with the help of time preference for weighting content similarity, our algorithm shows better performance than un-weighted one, which is content based filtering (CBF method). This could demonstrate that our proposed time preference has promising effects over the task of citation recommendation. Besides, the baselines which define the time related functions directly by prior knowledge didn't work well here. Such mechanism shows limitations in this study. For the baseline which applies dynamic topic model, its performance is not good enough to compare with other methods. The possible reason could be the improper topic number that selected ahead which might lead to the poor representation learning of documents. For WHIN-CSL model, although WHIN-CSL_4 behaves the best among all different models, we need to tune parameters to obtain such results. For WHIN-CSL_1, WHIN-CSL_2 and WHIN-CSL_3, they all get worse performance than the other models.

Table 4 Evaluation metrics of recommendation methods using DBLP dataset

| Metrics / Method | MAP | NDCG | MMR | P@30 | R@30 |
| --- | --- | --- | --- | --- | --- |
| CBF | 0.0014 | 0.0112 | 0.0356 | 0.0070 | 0.0057 |
| DTM | 0.0117 | 0.0691 | 0.1469 | 0.0383 | 0.0325 |
| Citerank | 0.0015 | 0.0116 | 0.0388 | 0.0075 | 0.0062 |
| Freshness | 0.0009 | 0.0097 | 0.0213 | 0.0052 | 0.0043 |
| Preference | 0.0012 | 0.0106 | 0.0309 | 0.0059 | 0.0048 |
| WHIN-CSL_5 | 0.0025 | 0.0190 | 0.0581 | 0.0119 | 0.0098 |
| WHIN-CSL_6 | 0.0030 | 0.0233 | 0.0664 | 0.0143 | 0.0117 |
| Our model | 0.0015 | 0.0117 | 0.0407 | 0.0071 | 0.0059 |

| Metrics / Method | MAP@30 | MAP@100 | NDCG@30 | NDCG@100 |
| --- | --- | --- | --- | --- |
| CBF | 0.0011 | 0.0014 | 0.0079 | 0.0112 |
| DTM | 0.0077 | 0.0117 | 0.0420 | 0.0691 |
| Citerank | 0.0013 | 0.0015 | 0.0085 | 0.0116 |
| Freshness | 0.0006 | 0.0009 | 0.0051 | 0.0097 |
| Preference | 0.0010 | 0.0012 | 0.0066 | 0.0106 |
| WHIN-CSL_5 | 0.0020 | 0.0025 | 0.0132 | 0.0190 |
| WHIN-CSL_6 | 0.0024 | 0.0030 | 0.0156 | 0.0233 |
| Our model | 0.0013 | 0.0015 | 0.0084 | 0.0117 |

Table 4 shows evaluation metrics using DBLP dataset. For the model of WHIN-CSL, conditional probability contains abstract-abstract similarity, paper-paper vertex similarity and paper-author vertexes similarity. For WHIN-CSL_5, $w_1$ is set to be 0.8 and $w_2$ is set to be 0.1. For WHIN-CSL_6, $w_1$ is set to be 0.7 and $w_2$ is set to be 0.2. Experimental results of our model are obtained when $k = 180$, where k is paper number of query's nearest neighbors and MRR obtains the highest value at this point (There are some points that obtaining the highest MAP at the same time, so we changed to MRR as the benchmark for DBLP dataset). When using the DBLP data, our algorithm still shows better performance than un-weighted one (CBF). Dynamic topic model behaves the best among all different models. Since it behaves the worst using PubMed data, such model is not robust. By adding the similarity between author and paper, WHIN-CSL achieves good performance. However, it requires the complete data of author information, which is hard in real application scenario.

**4.2.2 Performance when adding more neighbor papers**

From Figure 5 to Figure 7, we showed the trend of mean cross entropy and mean reciprocal rank when increasing the neighbor papers number to infer citation network embedding of user query. It can be seen that, there is a clear decreasing trend for cross entropy (Figure 5) when predicting the time preference of test data and evaluation of citation recommendation is getting better at the same time. Apparently, more neighbor papers can add the possibility of enriching the representation of user query in some certain degree. When selecting more similar papers, such patterns are getting stronger to show evidence of potential citations, which is also referring to the citing preference over different time slices. Figure 6 to Figure 7 show that increasing neighbor papers will not certainly bring strong improvement, but there would a trend for model to get better performance. And the model could behave well in at some point (k=100 in Figure 6 and k=180 in Figure 7). This can explain that why many studies are attempting to investigate proper parameters in their experiments.

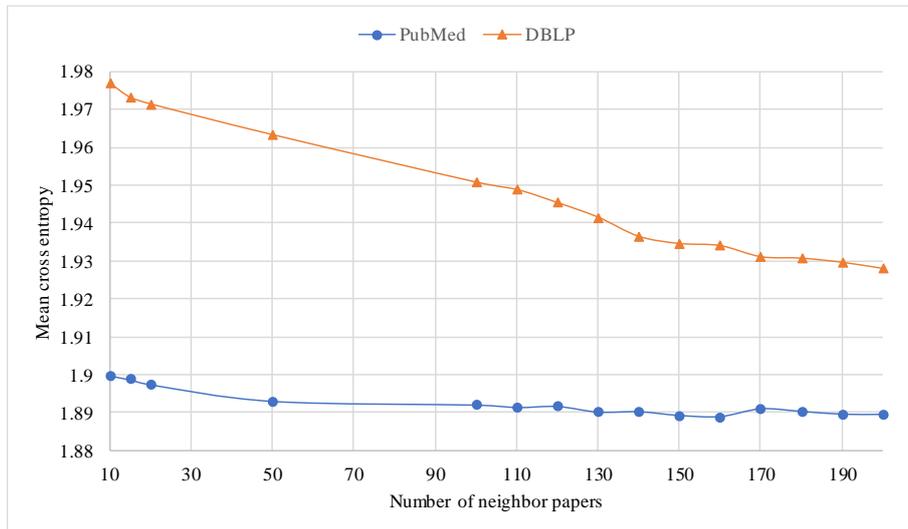

Figure 5: Mean cross entropy when increasing neighbor papers

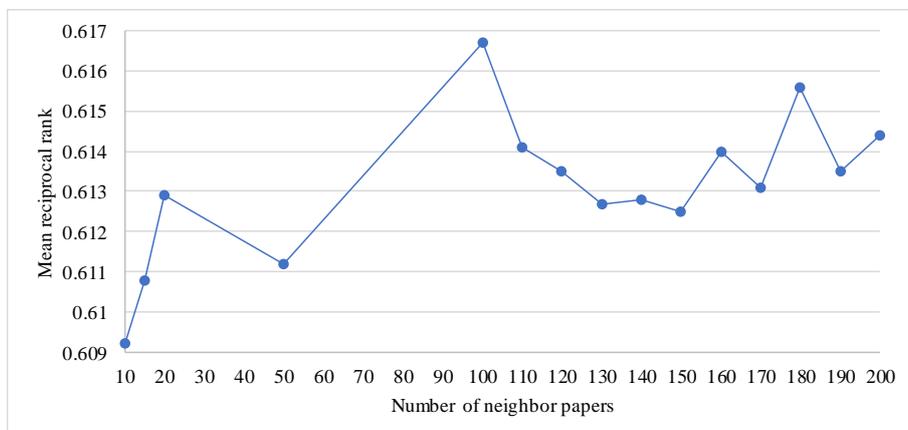

Figure 6: MMR when increasing neighbor papers using PubMed data

According to the experimental results, we can find that by adding citing time preference, the content-based model can get improved. Compared with other baselines, our method will not be affected a lot by parameter setting such topic numbers in DTM method. It also doesn't require full information of authors which are needed in WHIN-CSL model.

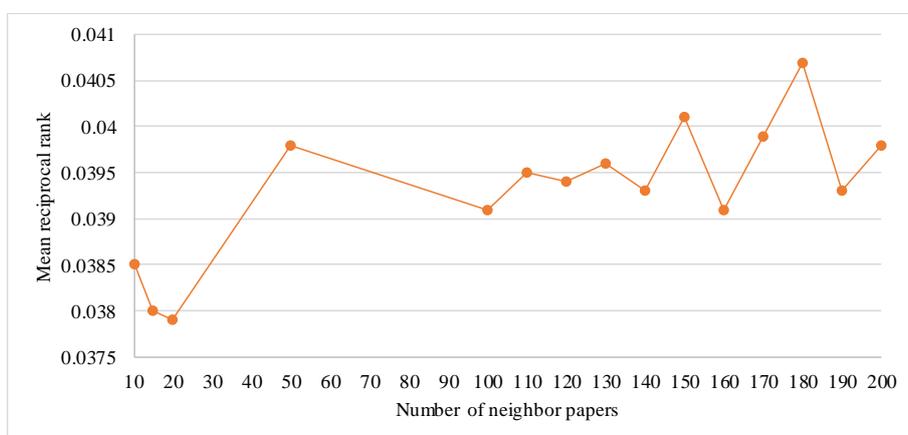

Figure 7: MMR when increasing neighbor papers using DBLP data

### 4.2.3 Performance when predicting preference with different distribution

Except analyzing the effects of neighbor paper numbers, we also want to look at the relationship between the prediction performance and the distribution of real time preference. Then, we calculate the standard deviation of real time preference and draw Figure 8 and Figure 9.

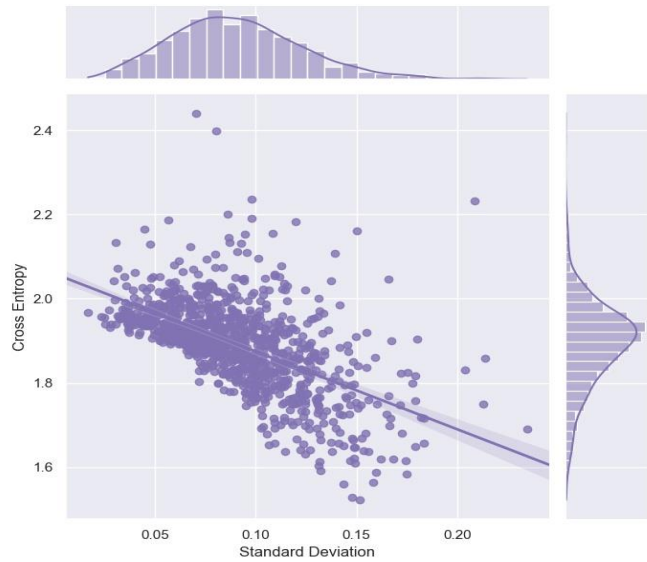

Figure 8: Cross entropy of prediction and standard deviation of real time preference using PubMed data

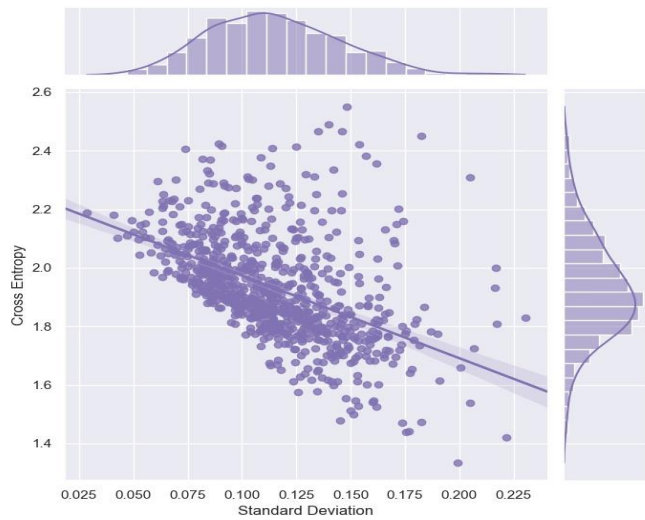

Figure 9: Cross entropy of prediction and standard deviation of real time preference using DBLP data

As we can see in Figure 8, most standard deviation values of real user preference range from 0.05 and 0.10, and most cross entropy of predicted preference range from 1.8 to 2.0. In Figure 9, most standard deviation values of real user preference range from 0.075 and 0.125, and most cross entropy of predicted preference range from 1.8 to 2.0. It can be found that when the real time preference has a low standard deviation (preference distribution is relatively homogeneous compared with others); the corresponding cross entropy value is relatively higher compared with the real time preference with a high standard deviation. It reflects that our model shows poorer performance when we predict time preference which has a homogeneous distribution in real. Better models should be constructed in the future for this kind of dataset.

**4.2.4 Analysis of recommended list**

To better understand how time preference improve candidate citation list, we compare our model with the content-based filtering method by displaying three different queries and their corresponding recommended list (the first 10 recommended true citations) obtained from PubMed dataset. Paper ID and real citing times of all citations are given here in Table 5, Table 6 and Table 7.

From the real citing times and the research problems described in query 22775499[9] and candidate citation 21048124[10], we can find that query paper published in 2013 conducts experiments based on the empirical evidence provided in candidate citation 21048124 published in 2010. This citation contains more detailed information about related methodology and conclusions, compared with another candidate citation 12722974[11] published in 2003. After utilizing the time preference learned from our model to weighting the CBF results, we are able to bring the truly related papers to the front of recommended list.

The same improvement is displayed in Table 4 as well. Query 20808844[12] is a paper published in 2010 named with "Structural alterations in a component of cytochrome c oxidase and molecular evolution of pathogenic Neisseria in humans". Candidate citation 18757819[13] is a paper published in 2008 titled with "Roles of c-type cytochromes in respiration in Neisseria meningitidis". These two papers are all related with cytochromes and Neisseria. For other two candidate citations 15335669[14] published in 1993 and 9157250[15] published in 1997, they are only cited for once in the query paper. However, due to the high content similarity, they are ranked ahead of citation 18757819 which is cited for 13 times. True time preference predicted for query 20808844 is (0.127, 0.225, 0.085, 0.113, 0.155, 0.282, 0.014) and the predicted one is (0.066, 0.107, 0.135, 0.148, 0.170, 0.211, 0.164). As we can see, the user has preference on the 6$^{th}$ time slice (2008-2009), and the predicted one also match this condition, then citation 18757819 can come first after weighting. Similarly, citation 19251655[16] published in 2009 also has changed its position in the recommended list.

---

[9] Available at: https://www.ncbi.nlm.nih.gov/pubmed/?term=22775499%5Buid%5D
[10] Available at: https://www.ncbi.nlm.nih.gov/pubmed/?term=21048124%5Buid%5D
[11] Available at: https://www.ncbi.nlm.nih.gov/pubmed/?term=12722974%5Buid%5D
[12] Available at: https://www.ncbi.nlm.nih.gov/pubmed/?term=20808844%5Buid%5D
[13] Available at: https://www.ncbi.nlm.nih.gov/pubmed/?term=18757819%5Buid%5D
[14] Available at: https://www.ncbi.nlm.nih.gov/pubmed/?term=15335669%5Buid%5D
[15] Available at: https://www.ncbi.nlm.nih.gov/pubmed/?term=9157250%5Buid%5D
[16] Available at: https://www.ncbi.nlm.nih.gov/pubmed/?term=19251655%5Buid%5D

Table 5 Top 10 recommended list of Query 22775499

| Query | Our model | | CBF | |
|---|---|---|---|---|
| | Citation | Citing times | Citation | Citing times |
| 22775499 | 22470333 | 16 | 18957228 | 16 |
| | 18957228 | 16 | 22470333 | 16 |
| | 21048124 | 22 | 12722974 | 3 |
| | 12722974 | 3 | 10513581 | 8 |
| | 15631589 | 19 | 15631589 | 19 |
| | 10513581 | 8 | 10050854 | 6 |
| | 19778516 | 3 | 21048124 | 22 |
| | 20702863 | 2 | 8269040 | 2 |
| | 10050854 | 6 | 19778516 | 3 |
| | 8269040 | 2 | 16781044 | 2 |

Table 6 Top 10 recommended list of Query 20808844

| Query | Our model | | CBF | |
|---|---|---|---|---|
| | Citation | Citing times | Citation | Citing times |
| 20808844 | 18757819 | 13 | 15335669 | 1 |
| | 18618739 | 1 | 9157250 | 1 |
| | 9157250 | 1 | 18757819 | 13 |
| | 15335669 | 1 | 18618739 | 1 |
| | 19251655 | 10 | 10948150 | 1 |
| | 10948150 | 1 | 10844690 | 1 |
| | 18353797 | 1 | 9413436 | 1 |
| | 19252222 | 1 | 19251655 | 10 |
| | 18403712 | 1 | 18353797 | 1 |
| | 10844690 | 1 | 19252222 | 1 |

Table 7 Top 10 recommended list of Query 22024163

| Query | Our model | | CBF | |
|---|---|---|---|---|
| | Citation | Citing times | Citation | Citing times |
| 22024163 | 20624899 | 2 | 20624899 | 2 |
| | 20921135 | 5 | 20921135 | 5 |
| | 21587233 | 1 | 21587233 | 1 |
| | 12719470 | 5 | 20956539 | 3 |
| | 18426974 | 2 | 20705815 | 2 |
| | 20956539 | 3 | 18426974 | 2 |
| | 11050385 | 1 | 12719470 | 5 |
| | 17276342 | 10 | 18243099 | 2 |
| | 12408861 | 1 | 19951914 | 1 |
| | 20705815 | 2 | 11050385 | 1 |

In Table 5, we can find that our predicted time preference can bring the candidate citation 17276342[17] which is cited 10 times into Top 10 recommended list while the content-based filtering algorithm failed to do this. To sum up, our predicted time preference can re-rank those which are topic-related and should be cited due to the chronological nature in some cases.

## 5 Conclusion and Future Work

Normally, the primary function of citation recommendation is to locate papers that are relevant to the user's information need. In this paper, we explored the hidden time preference of people's citing behavior and applied it to re-rank recommended list obtained by content-based filtering. Although the experimental results don't show significant superiority over other time-based models. By studying examples carefully, we find that, after utilizing the time preference to weighting the CBF results, our recommender is able to bring those truly related papers to the front of recommended list. Besides, such time preference can be added into any other citation recommendation framework for further refinement.

The methodological limitations of this work are as follows. First, although the node embedding is trained using citation graph without query data. The citation graph shouldn't contain any papers published later than queries. In the future work, we could select papers published in the latest time slice to be queries for both training and testing set, which will be more similar with application scenario. Also, other strategies should be conducted to make full use of the time preference. Second, better deep learning models should be built to see if there is still space for improvement of predictor performance. Third, we only use abstract content to represent each paper which definitely limits representation learning of content vectors, full text data should be taken into consideration lately and more bibliographic data should be tested in the future to validate our model fully.

## Acknowledgment

This work is supported by the National Natural Science Foundation of China (No. 72074113), Science Fund for Creative Research Group of the National Natural Science Foundation of China (No. 71921002) and Major Projects of National Social Science Fund (No. 16ZDA224).

---

[17] Available at: https://www.ncbi.nlm.nih.gov/pubmed/?term=17276342%5Buid%5D